\documentclass[aps,prb,twocolumn,unsortedaddress,floatfix,nofootinbib,superscriptaddress]{revtex4-2}
\usepackage{amsthm}
\usepackage{amsfonts}
\usepackage{siunitx}
\usepackage{amsmath}
\usepackage{amssymb}
\usepackage{graphicx}
\usepackage{verbatim}
\usepackage[colorlinks]{hyperref}
\usepackage{tikz}
\usepackage{braket}
\usepackage{xcolor}

\makeatletter
\def\@ssect@ltx#1#2#3#4#5#6[#7]#8{%
  \def\H@svsec{\phantomsection}%
  \@tempskipa #5\relax
  \@ifdim{\@tempskipa>\z@}{%
    \begingroup
      \interlinepenalty \@M
      #6{%
       \@ifundefined{@hangfroms@#1}{\@hang@froms}{\csname @hangfroms@#1\endcsname}%
       {\hskip#3\relax\H@svsec}{#8}%
      }%
      \@@par
    \endgroup
    \@ifundefined{#1smark}{\@gobble}{\csname #1smark\endcsname}{#7}%
  }{%
    \def\@svsechd{%
      #6{%
       \@ifundefined{@runin@tos@#1}{\@runin@tos}{\csname @runin@tos@#1\endcsname}%
       {\hskip#3\relax\H@svsec}{#8}%
      }%
      \@ifundefined{#1smark}{\@gobble}{\csname #1smark\endcsname}{#7}%
      \addcontentsline{toc}{#1}{\protect\numberline{}#8}%
    }%
  }%
  \@xsect{#5}%
}%
\makeatother

\definecolor{linkcolor}{RGB}{0,83,166}
\hypersetup{
  colorlinks = true,
  allcolors = {linkcolor}
}

\begin{document}
\newcommand{\mytitle}{Multi-objective optimization by quantum annealing}

\author{Andrew D. King}
\affiliation{D-Wave Quantum Inc., 3033 Beta Ave., Burnaby, BC, Canada}

\title{\mytitle}

\date{\today}
\begin{abstract}An important task in multi-objective optimization is generating the Pareto front---the set of all Pareto-optimal compromises among multiple objective functions applied to the same set of variables.  Since this task can be computationally intensive even for small problems, it is a natural target for quantum optimization.  Indeed, this problem was recently approached using the quantum approximate optimization algorithm (QAOA) on an IBM gate-model processor~\cite{kotil_quantum_2025}.  Here we compare these QAOA results with quantum annealing on the same two input problems, using the same methodology.  We find that quantum annealing vastly outperforms not just QAOA run on the IBM processor, but all classical and quantum methods analyzed in the previous study.  On the harder problem, quantum annealing improves upon the best known Pareto front.  This small study reinforces the promise of quantum annealing in multi-objective optimization.
\end{abstract}

\maketitle
\def\title#1{\gdef\@title{#1}\gdef\THETITLE{#1}}

\section{Introduction}

In multi-objective optimization (MOO), one must simultaneously consider the priorities of multiple stakeholders.  Finding the set of Pareto-optimal compromises, in which we cannot improve one objective without degrading another, can be enormously difficult even when each individual objective function can be optimized easily~\cite{gunantara_review_2018}.  In particular, since this can apply to unconstrained binary problems, MOO is an attractive target for quantum optimization; in this study we consider the two most popular quantum optimization approaches.

The first is the quantum approximate optimization algorithm (QAOA)~\cite{farhi_quantum_2014,blekos_review_2024}, in which multiple layers (in this study, $p=6$) of mixer Hamiltonians and objective Hamiltonians are alternatingly applied to a quantum state.  The second is quantum annealing (QA)~\cite{kadowaki_quantum_1998,johnson_quantum_2011,rajak_quantum_2022,king_quantum_2023}, in which quantum fluctuations are attenuated, guiding an initial state through a quantum phase transition into a low-energy state of the classical target Hamiltonian---several recent studies have already reported promising results in applying QA to MOO \cite{aguilera_multiobjective_2024,schworm_multiobjective_2024,kuo_quantum_2025,wang_quantum_2025}.  QAOA and QA are related, falling into the same general framework of quantum optimization through the application of varying Hamiltonians~\cite{brady_optimal_2021,miessen_benchmarking_2024}.  However, the weight of evidence~\cite{pelofske_shortdepth_2024,mcgeoch_comment_2024}, including a comparison of the two on identical hardware~\cite{ebadi_quantum_2022}, suggests that QA is a more effective means of optimization on current quantum processors.  Nonetheless, QAOA, which has theoretical approximation guarantees in an ideal fault-tolerant QPU, has a comfortable place in the gate-model orthodoxy.

\begin{figure*}
  \includegraphics[width=17cm]{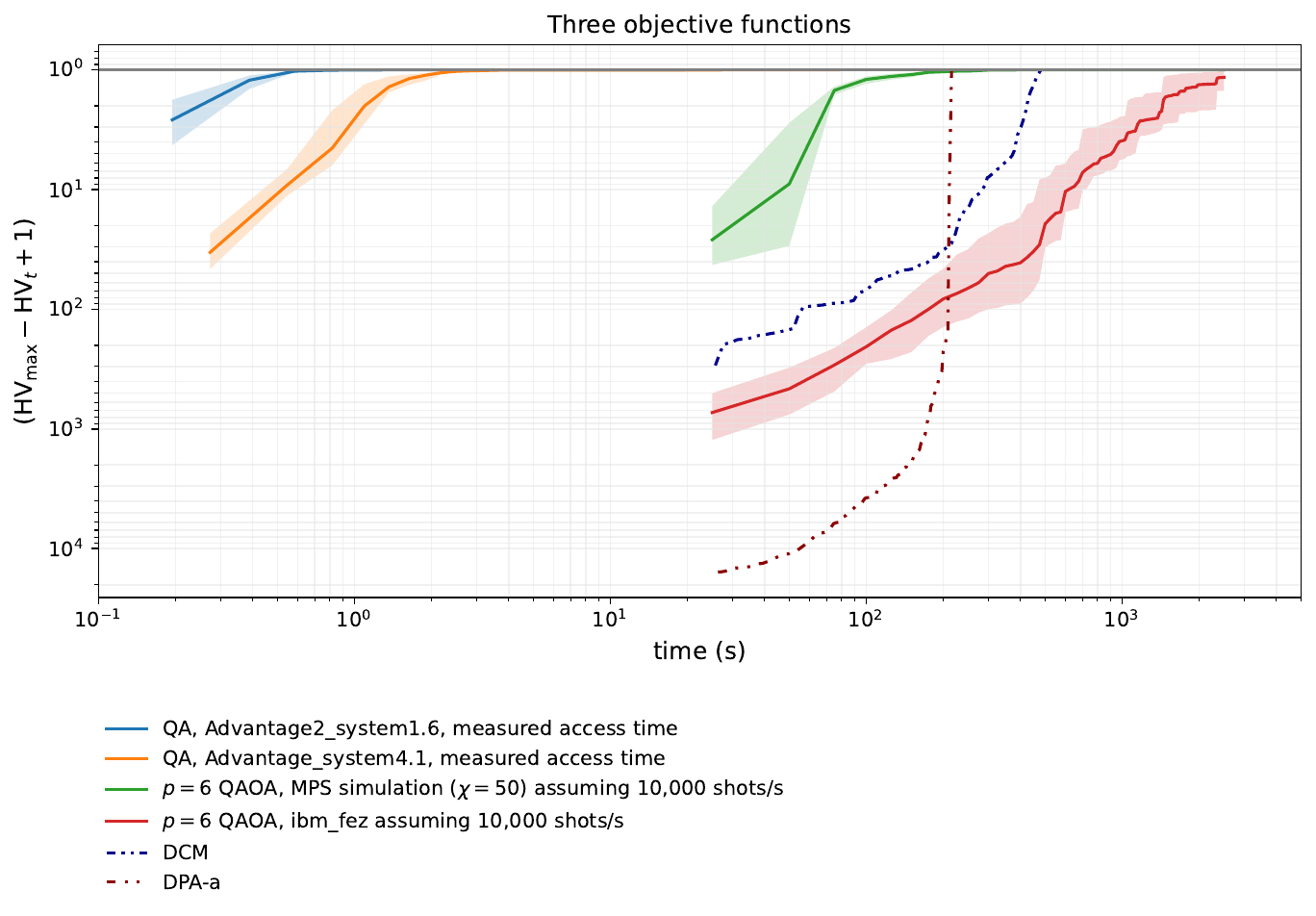}
\caption{{\bf Three-objective results}.  The figure of merit, $\text{HV}_\text{max}-\text{HV}+1$, is plotted for various classical and quantum approaches.  For sampling-based approaches, i.e.~QA and QAOA, experiments are run five times, with the mean indicated by a line and the shaded region indicating the best and worst among five runs.  Both QA systems found the same optimal solution as Ref.~\cite{kotil_quantum_2025}, with the Pareto front made up of 2067 non-dominated points.  All results other than those for QA are taken from Ref.~\cite{kotil_quantum_2025}.}\label{fig:1}
\end{figure*}

\section{Multi-objective weighted maximum-cut and QA}

A recent study of QAOA applied to multi-objective weighted maximum-cut problems~\cite{kotil_quantum_2025} found promising results, namely that noise-free matrix-product-state (MPS) simulations of QAOA could outperform classical approaches.  Here, we apply an identical methodology to solve the same MOO problems, but with a real QA system used in place of QAOA (both MPS-simulated QAOA and QAOA run on an actual quantum processor).

Weighted maximum-cut can be expressed as an Ising optimization problem---the native language of D-Wave quantum annealing processors---and the heavy-hex qubit connectivity graph of IBM quantum processors fits into that of D-Wave quantum annealing processors as a subgraph.  The task of replacing QAOA with QA in the workflow of Ref.~\cite{kotil_quantum_2025} is therefore straightforward.

Our aim is to take a graph $G$ on $N$ vertices with edge set $E$, and to maximize some combination of the $M$ objective functions $\{F_k\}_{k=1}^M$ defined as
\begin{equation}
F_k(s) = -\sum_{u,v\in E}s_is_jJ_{i,j,k}
  \end{equation}
  where $s$ is a vector of binary Ising variables in $\{-1,1\}^{N}$, and $J_{i,j,k}$ are edge weights.  In this case, edge weights take independent random Gaussian values, and the graph $G$ is a heavy-hex graph on $N=42$ nodes with 46 edges.

  To determine the Pareto front with QAOA, the approach taken by Ref.~\cite{kotil_quantum_2025} is the following: For many ($5000$ for $M=3$ or 20,000 for $M=4$) random relative weightings $\{c_k\}_{k=1}^M$ with $c_k\geq 0$ and $\sum_{k=1}^Mc_k=1$, draw $5000$ approximately optimal samples from the weighted objective function $\sum_{k=1}^Mc_kF_k$.  This gives a set $S$ of states.  The set of non-dominated states (states $s$ of $S$ such that for every other $s'$ in $S$, there is some $i$ for which $F_i(s)>F_i(s')$) forms an approximation of the Pareto front.

  In QA, we take the same approach, but draw only $1000$ samples for each $c$ vector.  Using the {\texttt{Advantage2\_system1.6}} solver, we can pack 96 disjoint copies of $G$ into the qubit connectivity graph, allowing us to sample from $96$ $c$ vectors in parallel (to probe the importance of QA sample quality, we also present data from the previous-generation \texttt{Advantage\_system4.1} solver, which is noisier but can sample from $114$ $c$ vectors in parallel).  We run anneals of duration $\SI{1}{\micro s}$, meaning that the overall duty cycle of the QA processor is dominated by readout time ($\SI{98}{\micro s}$ and $\SI{235}{\micro s}$ per sample, respectively, for {\texttt{Advantage2\_system1.6}} and  \texttt{Advantage\_system4.1}).  We report total QPU access time, including programming and readout, which totals roughly $\SI{0.2}{s}$ for a $1000$-shot QPU call yielding 96,000 or 114,000 42-qubit samples.  We auto-scale each QPU call to maximize coupling energies within the programmable range $J_{i,j}\in [-2,1]$, but do not use spin-reversal transformations to further boost the energy scale, though this was previously shown to be effective in heavy-hex spin glasses~\cite{mcgeoch_comment_2024}.

  Following Ref.~\cite{kotil_quantum_2025}, we use the hypervolume (HV) of non-dominated points as a figure of merit.  This is the volume of the union of all hyperrectangles in $M$-dimensional space between a common reference point $r$ and the $M$-dimensional vector $\{F_k\}_{k=1}^M$ for each non-dominated sample.  The reference point $r$ is taken as the vector of {\it minimum} values of each $F_k$---this serves as an arbitrary point that must be dominated by any point in the Pareto front. The hypervolume of the entire Pareto front (assuming we have found the entire front) is denoted $\text{HV}_\text{max}$.  For a given set of samples, both the non-dominated subset and hypervolume can be time consuming to compute; we use the algorithm in the {\texttt{moocore}} Python package~\cite{lopez-ibanez_exploratory_2010, moocore}.

\begin{figure*}
  \includegraphics[width=17cm]{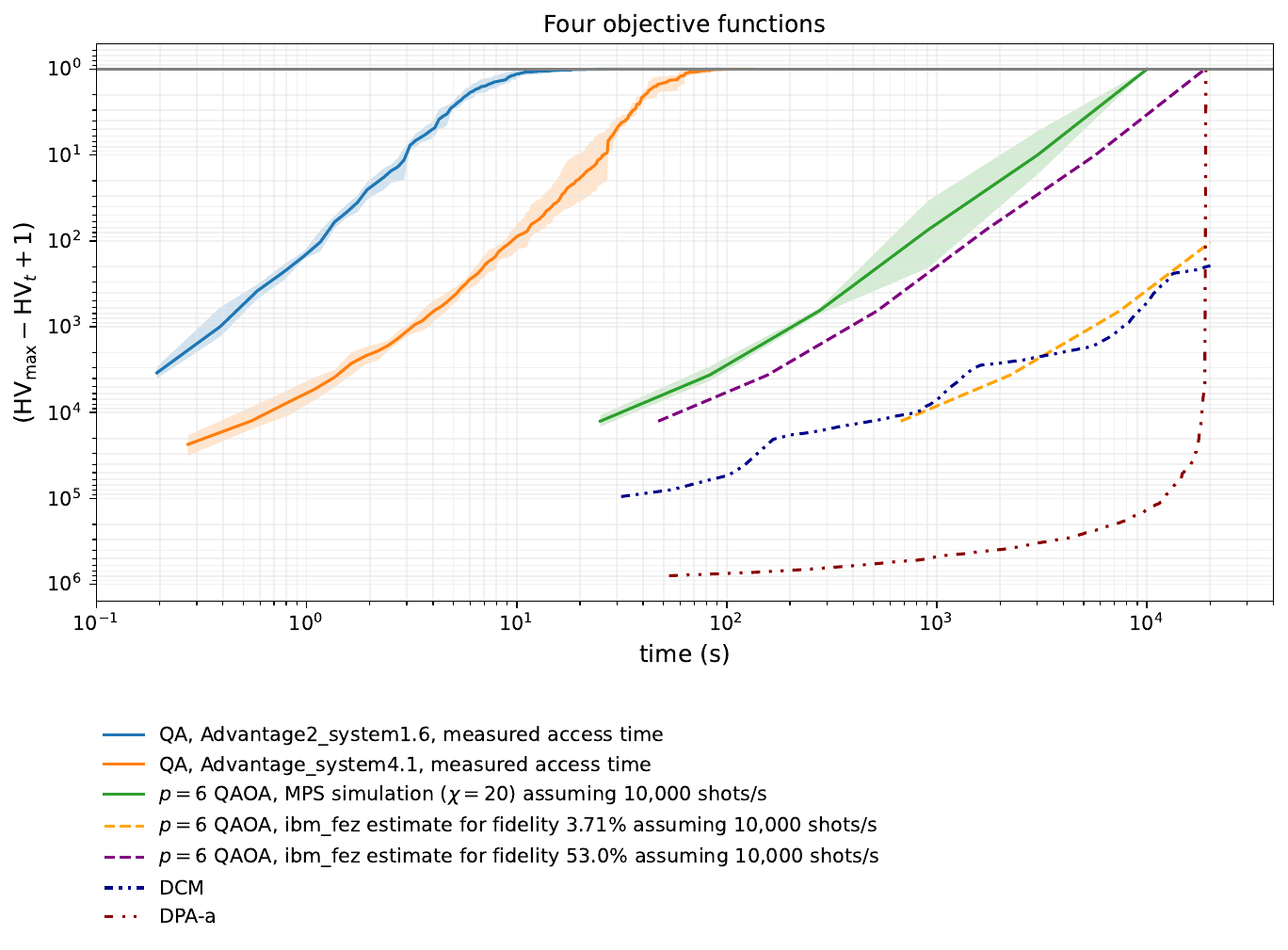}
  \caption{{\bf Four-objective results}.  Data is as in Fig.~\ref{fig:1}, but no results are available for \texttt{ibm\_fez}.  Estimates are derived from MPS results assuming circuit fidelities of 53\% and 3.71\% (the latter being an estimate for current systems).  QA found a better Pareto front than all other approaches, consisting of 30,419 non-dominated points.  All results other than those for QA are taken from Ref.~\cite{kotil_quantum_2025}.}\label{fig:2}
\end{figure*}

  \section{Results}
  
  Ref.~\cite{kotil_quantum_2025} presented results on two problems, one with three objective functions and one with four; their results are presented along with ours in Figs.~\ref{fig:1} and \ref{fig:2} respectively.  Ref.~\cite{kotil_quantum_2025} uses two methods to generate samples for the aforementioned approach to building up the Pareto front: actual QAOA experiments using the \texttt{ibm\_fez} processor, and MPS simulations of QAOA circuits with varying bond dimension $\chi$ ($\chi$ is used to tune the tradeoff between computational resources and sample quality).  For both, they plot data assuming a rate of 10,000 shots per second, although this is based on inter-circuit delay and not actually measured across the total IBM QPU duty cycle.  MPS is used as a simulator for a hypothetical noise-free quantum processor.  Shaded regions indicate the range of performance across five independent repetitions.

  As seen in Fig.~\ref{fig:1}, QA ({\texttt{Advantage2\_system1.6}}) approaches the optimal hypervolume in just three QPU calls, and in the median case finds the optimal Pareto front---consisting of 2067 non-dominated points---in under two seconds; this is over 100 times faster than the hypothetical MPS simulation and 1000 times faster than the QAOA experiments on \texttt{ibm\_fez} (\texttt{ibm\_fez} experiments do not reach optimality).
  
  Also included for comparison are two fully classical algorithms, DCM and DPA-a, which are integer-based algorithms run on discretized weights, and therefore do not necessarily give optimal Pareto fronts for the continuous Gaussian weights.  Indeed, in the four-objective problem, whose hypervolumes are shown in Fig.~\ref{fig:2}, QA found 30,419 non-dominated points that include the 30,409 points reported in Ref.~\cite{kotil_quantum_2025}.  Thus none of the solvers other than QA find the optimal Pareto front; we believe this front of 30,419 points to be optimal since QA found it routinely and repeatedly.  In the median case, this took 203 QPU calls for {\texttt{Advantage2\_system1.6}}, or under 20 million 42-qubit samples, compared with the 100 million samples taken by the MPS QAOA simulation.  The approach to optimality is roughly 1000 times faster for QA than for the hypothetical MPS QAOA simulation, when MPS is assumed to provide 10,000 shots per second.

  \section{Conclusion}

  We have presented a simple reproduction of a quantum multi-objective optimization workflow, in which replacing QAOA with quantum annealing leads to a speedup of multiple orders of magnitude.  This involved no advanced parameter tuning and we provide complete source code and data~\cite{QuantumAnnealingMOO}.  Our results are consistent with previous observations that QAOA is not competitive with QA in binary optimization tasks.

\bibliography{paper}

\end{document}